\documentclass[reprint, aps, pra, floatfix, superscriptaddress, nofootinbib]{revtex4-1}
\usepackage{graphicx}    % needed for figures
\usepackage{dcolumn}    % needed for some tables
\usepackage{bm}             % for math
\usepackage{amssymb}   % for math
\usepackage{amsmath}    % for multiline equation
\usepackage{amsthm}
\usepackage{mathrsfs}    % for mathscr{}
\usepackage{commath}   % for absolute value
\usepackage{subfigure}    % for subfigure
\usepackage{braket}
\usepackage{natbib}
\usepackage{float}
\usepackage{color}
\usepackage{siunitx}
\usepackage[colorlinks=true, allcolors=blue]{hyperref}
\usepackage{hyphenat}
\usepackage{notes2bib}

\graphicspath{{Figures/}}

\begin{document}

\title{Quantum transduction with microwave and optical entanglement}

\author{Changchun Zhong}
\email{zhong.changchun@uchicago.edu}
\affiliation{Pritzker School of Molecular Engineering, University of Chicago, Chicago, IL 60637, USA}

% \author{??}
% \affiliation{Pritzker School of Molecular Engineering, University of Chicago, Chicago, IL 60637, USA}

\author{Xu Han}

\affiliation{Center for Nanoscale Materials, Argonne National Laboratory, Argonne, Illinois 60439, USA}

% \author{??}
% \affiliation{Center for Nanoscale Materials, Argonne National Laboratory, Argonne, Illinois 60439, USA}

\author{Liang Jiang}
\affiliation{Pritzker School of Molecular Engineering, University of Chicago, Chicago, IL 60637, USA}

\date{\today}

\begin{abstract}
Quantum transduction refers to the coherent conversion between microwave and optical states, which can be achieved by quantum teleportation if given high fidelity microwave-optical entanglement, namely entanglement-based quantum transduction. Reliable microwave-optical entanglement can be generated using various platforms. In this paper, we base the discussion on piezo-optomechanical system and make the teleportation induced conversion scheme more concrete in the framework of quantum channel theory. By comparing the quantum capacity between the entanglement-based conversion channel and the traditional direct quantum transduction channel, we show entanglement-based scheme indeed admits a positive transduction rate when the direct quantum transduction has zero quantum capacity. Given two piezo-optomechanical systems, we also investigate the generation of microwave-microwave entanglement from entanglement swapping within continuous variable and discrete variable settings, showing the potentials of directly connecting microwave quantum processor by microwave-microwave quantum teleportation.

\end{abstract}

\maketitle

\section{Introduction}

Distant microwave quantum processors, connected by efficient optical quantum channels, form an important design of quantum network \cite{Cirac1997,Kimble2008}. This design is appealing since it tries to combine two very different and important fields: 1) the superconducting circuit known for its advantages including efficient quantum control, hardware scalability, etc  \cite{blais2021}; 2) the optical quantum channels for quantum information transmission with the feature of low communication loss, room temperature quantum coherence preserving, etc \cite{tittel1998,yin2017}. Since optical and microwave photons do not interact, to build up this quantum network, a quantum transducer which coherently convert quantum information between microwave and optical frequencies is indispensable. However, establishing such a quantum interface is extremely challenging to the current technology because the traditional direct quantum transduction (DQT), which linearly converts photons with beam-splitter-type coupling, requires both high coupling efficiency and small added noise \cite{andrews2014,vainsencher16,mirhosseini2020}. Currently, DQT is being actively studied with various physical platforms, e.g., electro-optomechanics \cite{Regal2011,Bochmann2013,Taylor2011,Barzanjeh2011,Wang2012,Tian2010,*Tian2012,*Tian2014,Midolo2018,Bagci2014,Winger2011,Pitanti2015}, electro-optics \cite{Tsang2010,*Tsang2011,Javerzac-Galy2016,Fan2018,fu2021}, quantum magnonics \cite{Hisatomi2016,zhu2020}, Rydberg atoms \cite{han2018}, etc. Although enormous progress has been made in the past several years for each platform, all of them are still below the level, only above which direct quantum state conversion is possible.

Quantum state conversion can alternatively be realized by entanglement-based quantum transduction (EQT), which first generates microwave-optical (MO) entanglement then completes the state conversion through quantum teleportation \cite{Barzanjeh2012,zhong2020}. Since a classical communication channel is used, EQT is expected to tolerate more noises and thus is less demanding in experiments. In fact, a series of recent studies already show the potential of high-fidelity MO entanglement generation based on the hybrid quantum systems in experimental feasible regime \cite{Barzanjeh2012,tian2013,zhong2020_,rueda2019}, which paves the way of quantum transduction in the near term.

In this paper, we compare the EQT and DQT schemes based on the platform of piezo-optomechanics. For DQT, we map out the system parameters where any quantum state conversion is impossible. While in the same parameter regime, we find the EQT still admits a finite quantum conversion rate, which is consistent with the result in quantum channel theory that a quantum channel with the assistance of classical communication could tolerate more noises \cite{weedbrook2021,garcia2009}. In addition, we discuss the generation of MO entanglement by entanglement swapping with the entangled MO sources generated from two piezo-optomechanical systems \cite{alessio2017,furusawa1998}. The microwave-microwave (MM) entanglement can be used to transmit microwave quantum information through teleportation induced channels, thus directly connecting distant microwave circuits. Our calculation shows that this induced transmission channel still has better feasibility than DQT, indicating a promising alternative scheme for realizing microwave quantum processor connections.

In the sections that follow, we first introduce the piezo-optomechanical system, based on which the DQT and EQT schemes are compared in the framework of quantum channel theory. The transduction scheme from entanglement swapping is studied in the end. Throughout the paper, the convention $\hbar=2$ is used for numerical calculations unless specified otherwise.

\section{Piezo-optomechanics}

\begin{figure}[t]
\centering
\includegraphics[width=\columnwidth]{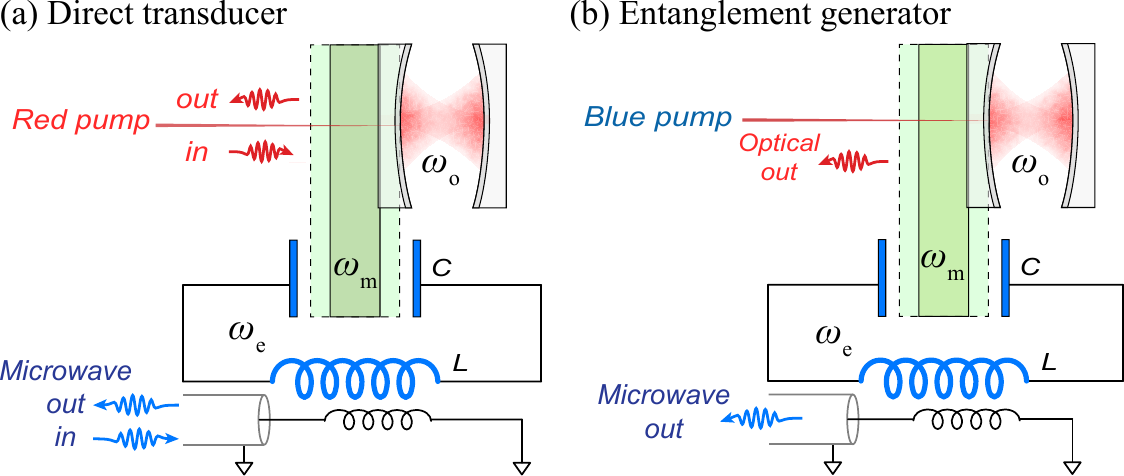}
\caption{Schematic figure for piezo-optomechanical system, used for (a) direct quantum transduction with a red detuned laser pump and for (b) entanglement generator with a blue detuned laser pump. \label{fig0}}
\end{figure}

We base our discussion on a piezo-optomechanical system, as shown schematically in Fig.~\ref{fig0}. The thickness mode of a mechanical resonator is on the one hand coupled to the microwave mode through piezo-electricity, and on the other hand coupled to the optical mode by scattering pressure \cite{xu2020}. Denote $\hat{a}$, $\hat{b}$ and $\hat{c}$ as the optical, mechanical, and microwave mode operators, respectively, and $\omega_{\mathrm{o}}$, $\omega_{\mathrm{m}}$, and $\omega_{\mathrm{e}}$ as the corresponding resonant frequencies. We use a laser with frequency $\omega_\text{L}$ to pump the optical mode and populate it with on average $\bar{n}_\text{o}$ photons. In the rotating frame of the laser, we can write down the linearized Hamiltonian of the system
\begin{equation}\label{hamil0}
\begin{split}
{\hat{H}}/{\hbar} &=  -\Delta_\mathrm{o} \hat{a}^\dagger\hat{a} + \omega_{\mathrm{m}} \hat{b}^\dagger \hat{b} + \omega_{\mathrm{e}} \hat{c}^\dagger \hat{c} - g_{\mathrm{em}} (\hat{b}^\dagger \hat{c} + \hat{b} \hat{c}^\dagger)\\
&\quad\, -g_{\mathrm{om,0}} \sqrt{\bar{n}_\mathrm{o}} (\hat{a}^\dagger + \hat{a}) (\hat{b}^\dagger + \hat{b}),
\end{split}
\end{equation}
where $\Delta_\text{o}=\omega_\mathrm{L}-\omega_\text{o}$, $g_\text{em}$ is the piezo-mechanical coupling strength and $g_\text{om,0}$ is the single photon scattering pressure coupling, which can be enhanced by the cavity photons. The enhanced coupling strength is denoted as $g_\text{om}=g_\text{om,0}\sqrt{\bar{n}_\text{o}}$. Generally, in the piezo-optomechanical system, the mechanical resonator is intrinsically coupled to a thermal bath with temperature around $10$ mili-Kelvin to $1$ Kelvin. For several GHz mechanical resonator, the thermal noise can be routinely maintained around $1$ \cite{mingrui19,xu2020}. Although this is already a remarkable progress experimentally, as shown later, to demonstrate the quantum state conversion in sub-photon level is still challenging. In the following discussions, we denote the intrinsic loss rate as $\kappa_\text{m}$ for mechanical mode and $\kappa_\text{e,i}$ for microwave. The microwave mode also has a coupling port with loss rate $\kappa_\text{e,c}$. On the optical side, we denote the optical coupling and intrinsic loss rate as $\kappa_\text{o,c}$ and $\kappa_\text{o,i}$. Note the optical mode typically has frequency in the THz regime, making the optical noise negligible even in room temperature, and thus is neglected in later discussions.

\section{Direct quantum transduction}

\subsection{Piezo-optomechanical system for direct quantum transduction}

Direct quantum transduction linearly converts quantum states coherently between microwave and optical regime with the help of beam splitter type interaction. This interaction can be generally obtained in many hybrid quantum systems \cite{zhu2020,andrews2014,Bochmann2013}. For piezo-optomechanics,
if we pump the optical mode in the red side band with $\Delta_\text{o}<0$, the Hamiltonian Eq.~\ref{hamil0} can be further simplified with rotating wave approximation
\begin{equation}
\begin{split}
    \hat{H}/\hbar=&-\Delta_\text{o}\hat{a}^\dagger\hat{a}+\omega_\text{m}\hat{b}^\dagger\hat{b}+\omega_\text{e}\hat{c}^\dagger\hat{c}+g_\text{om}(\hat{a}^\dagger\hat{b}+\hat{a}\hat{b}^\dagger)\\
    &+g_\text{em}(\hat{b}^\dagger\hat{c}+\hat{b}\hat{c}^\dagger).
\end{split}    
\end{equation}
We see a beam splitter interaction is generated between the optical and mechanical modes, enabling the state conversion from optical to mechanical resonator and vice versa. The mechanical resonator further piezo-mechanically couples to the microwave mode and swaps the state between them, thus indirectly realizing bidirectional microwave-optical transduction. To quantify this conversion process in detail, we first write down the Heisenberg-Langevin equations for each mode and the input-output relations as
\begin{equation}
    \begin{split}
        \dot{\textbf{a}}&=\textbf{A}\textbf{a}+\textbf{B}\textbf{a}_\text{in}\\
        \textbf{a}_\textbf{out}&=\textbf{B}^\text{T}\textbf{a}-\textbf{a}_\text{in}
    \end{split}
\end{equation}
where we label the vectors $\textbf{a}=\{\hat{a},\hat{c},\hat{b}\}^\text{T}$, $\textbf{a}_\text{in}=\{\hat{a}_\text{in,c},\hat{a}_\text{in,i},\hat{c}_\text{in,c},\hat{c}_\text{in,i},\hat{b}_\text{in}\}^\text{T}$ and $\textbf{a}_\text{out}=\{\hat{a}_\text{out,c},\hat{a}_\text{out,i},\hat{c}_\text{out,c},\hat{c}_\text{out,i},\hat{b}_\text{out}\}^\text{T}$. The lower indexes ``in/out" indicate the input and output modes, while ``c/i" represent the coupling and intrinsic loss ports. The matrices 
\begin{equation}
    \textbf{A}=
    \begin{pmatrix}
    i\Delta_\text{o}-\frac{\kappa_\text{o}}{2}& 0 & -ig_\text{om}\\
    0 & -i\omega_\text{e}-\frac{\kappa_\text{e}}{2} & -ig_\text{em}\\
    -ig_\text{om} & -ig_\text{em} & -i\omega_\text{m}-\frac{\kappa_\text{m}}{2} 
    \end{pmatrix}
\end{equation}
and
\begin{equation}
    \textbf{B}=
    \begin{pmatrix}
    \sqrt{\kappa_\text{o,c}} &\sqrt{\kappa_\text{o,i}} & 0 & 0 & 0\\
    0 & 0 & \sqrt{\kappa_\text{e,c}} & \sqrt{\kappa_\text{e,i}} & 0 \\
    0 &0 &0& 0& \sqrt{\kappa_\text{m}}
    \end{pmatrix}.
\end{equation}
The above equation group can be solved in the frequency domain by taking the Fourier transform $\hat{o}[\omega]=\int dt\hat{o}(t)e^{i\omega t}$, where $\hat{o}$ denotes an arbitrary operator. Straightforwardly, the input and output modes are shown to be connected by the scattering relation
\begin{equation}\label{tlchannel}
    \textbf{a}_\text{out}[\omega]=\textbf{S}[\omega]\cdot\textbf{a}_\text{in}[\omega]
\end{equation}
where $\textbf{S}[\omega]=\textbf{B}^\text{T}(-i\omega \textbf{I}_3-\textbf{A})^{-1}\textbf{B}-\textbf{I}_5$. Based on the scattering matrix, we can identify the quantum transduction channel, e.g., with the on resonance condition ($\omega_\text{m}=\omega_\text{e}=-\Delta_\text{o}$), the microwave to optical conversion channel is can be written down as
\begin{equation}
    \hat{a}_\text{out,c}=\sqrt{\eta}\hat{c}_\text{in,c}+\sqrt{1-\eta}\hat{e}
\end{equation}
which is interpreted as a beam splitter mixing the input signal and the thermal noise. $\eta$ is the conversion efficiency
\begin{equation}\label{Cefficiency}
    \eta=\frac{4C_\text{om}C_\text{em}}{(1+C_\text{om}+C_\text{em})^2}\zeta_\text{o}\zeta_\text{e}.
\end{equation}
$\zeta_\text{o}=\kappa_\text{o,c}/\kappa_\text{o}$ and $\zeta_\text{e}=\kappa_\text{e,c}/\kappa_\text{e}$ are the extraction ratios and the system cooperativities are given by $C_\text{om}=4g_\text{om}^2/\kappa_\text{o}\kappa_\text{m}$ and $C_\text{em}=4g_\text{em}^2/\kappa_\text{e}\kappa_\text{m}$. Note this efficiency is obtained for $\omega=0$ which is optimal for weakly coupled system \cite{zhong2020_}. $\hat{e}$ is a noise input operator defined as
\begin{equation}
    \hat{e}=\frac{1}{\sqrt{1-\eta}}(S_{11}\hat{a}_\text{in,c}+S_{12}\hat{a}_\text{in,i}+S_{14}\hat{c}_\text{in,i}+S_{15}\hat{b}_\text{in}).
\end{equation}
If we ignore safely the optical noises, the average input noise photon is obtained
\begin{equation}
    n_e=\frac{1}{1-\eta}(|S_{14}|^2n_c+|S_{15}|^2n_b),
\end{equation}
where $|S_{14}|^2=\frac{4C_\text{em}C_\text{om}}{(1+C_\text{om}+C_\text{em})^2}\zeta_\text{o}(1-\zeta_\text{e})$ and $|S_{15}|^2=\frac{4C_\text{om}}{(1+C_\text{om}+C_\text{em})^2}\zeta_\text{o}$. In the piezo-optomechanical system, the mechanical mode and the microwave mode are intrinsically coupled to the same thermal bath with temperature $\mathcal{T}$, indicating $n_b=n_c=n_\text{th}\equiv(e^{{\hbar\omega_\text{m}/k_\text{B}\mathcal{T}}}-1)^{-1}$. Thus for finite bath temperature, the microwave-optical conversion is a Bosonic thermal loss channel with transmissivity $\eta$ and thermal noise $n_e$. We denote it as $\mathcal{N}(\eta,n_e)$ which maps an input state with covariance matrix $\mathbf{V}$ into $\mathbf{TVT}^\mathrm{T}+\mathbf{N}$, where $\mathbf{T}=\sqrt{\eta}\mathbf{I}_2$ and $\mathbf{N}=(1-\eta)(2n_e+1)\mathbf{I}_2$ (see the appendix for a brief review of the Bosonic channel representation). 

\begin{figure}[t]
\centering
\includegraphics[width=\columnwidth]{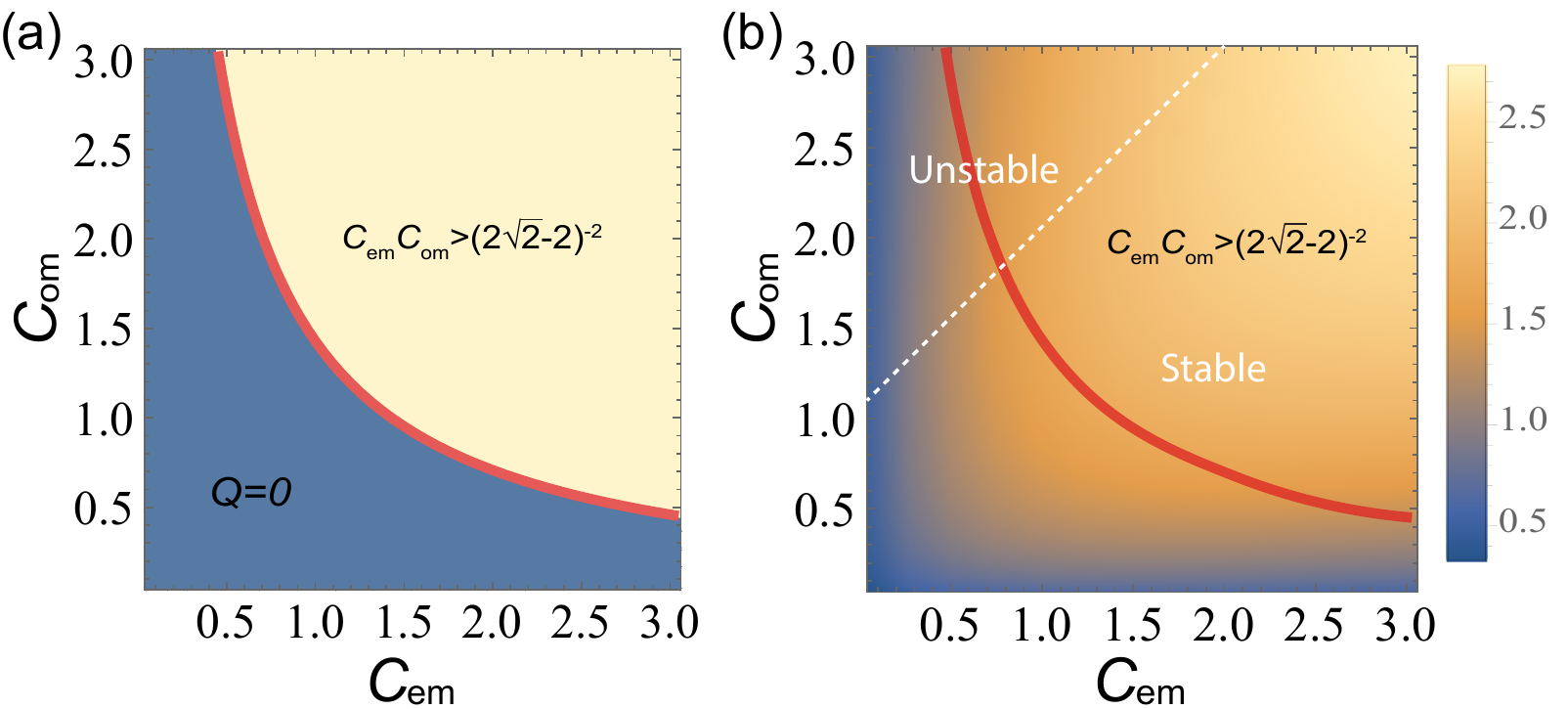}
\caption{(a) Direct quantum transduction behavior for piezo-optomechanical system, where the red line separate the parameter regime between the quantum capacity is sure to be zero and the quantum capacity can possibly be positive. For ideal case (zero thermal noise, unit extraction ratio), the red line becomes the exact quantum capacity threshold; (b) The capacity lower bound of entanglement-based transduction channel in the ideal case (zero thermal noise, unit extraction ratio), showing positive transduction rate in the regime where the direct transduction can only have zero capacity.  \label{fig1}}
\end{figure}

A channel is able to transmit quantum information as long as it has positive quantum channel capacity (see the appendix for a brief review). For many quantum channels including thermal loss channel, finding their exact expressions of quantum capacity is hard. Instead, we resort to the capacity lower bound to study the channel properties. The channel $\mathcal{N}(\eta,n_e)$ admits a capacity lower bound \cite{weedbrook2021}
\begin{equation}\label{clower}
    Q_\text{LB}^\mathcal{N}=\max\{0,\log_2\frac{\eta}{1-\eta}-g(n_e)\},
\end{equation}
where $g(x)\equiv(x+1)\log_2(x+1)-x\log_2x$. Interestingly, this bound is tight for so called pure loss channel ($n_e=0$). For pure loss channel, it is easy to get that $\eta=1/2$ is the threshold to have a positive channel capacity. Thus for thermal loss channel, it is necessary to have $\eta>1/2$ in order to get a positive quantum capacity, since the thermal noise generally degrades the channel. Using this necessary condition and the expression Eq.~\ref{Cefficiency}, we have (noticing $\zeta_\text{e}\zeta_\text{o}\le1$)
\begin{equation}
    C_\text{om}C_\text{em}>\left(\frac{1}{2\sqrt{2\zeta_\text{o}\zeta_\text{e}}-2}\right)^2\ge\frac{1}{(2\sqrt{2}-2)^2}
\end{equation}
as the least requirement of the system to have positive channel capacity, as shown in Fig.~\ref{fig1}(a). It is worth noting that this condition is not sufficient since the thermal noise as well as the non-unit extraction ratios could further degrade the channel behaviors. Currently, a great effort is being put on designing and improving the experimental devices. Although a huge progress has been made in the past decade \cite{Higginbotham2018,xu2020,vainsencher16,mirhosseini2020}, the required parameter regime for positive capacity is still hard to reach with the state-of-the-art technology. 

\section{Entanglement-based quantum transduction}

\subsection{MO entanglement from piezo-optomechanical system}

\begin{figure*}[t]
\centering
\includegraphics[width=\textwidth]{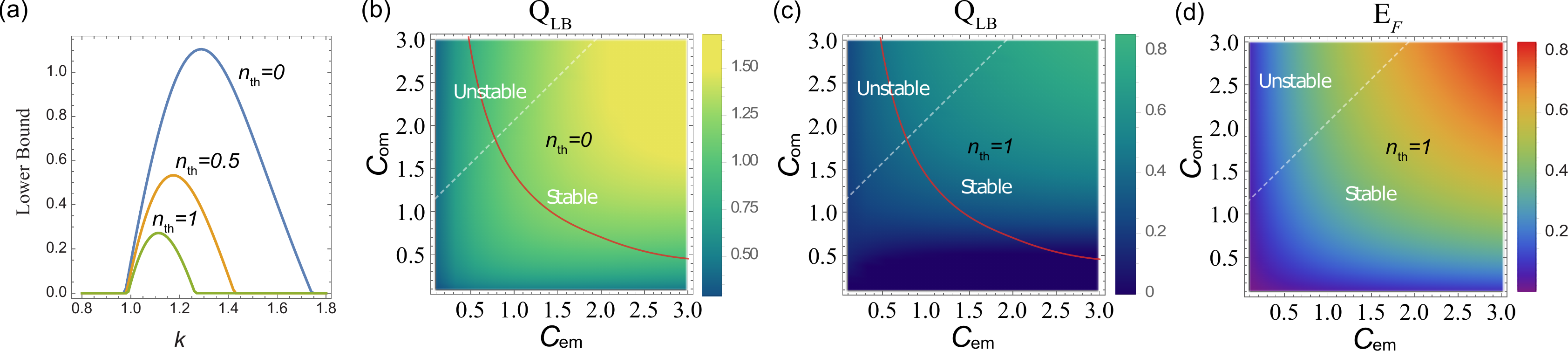}
\caption{(a) Quantum capacity lower bound of teleportation induced conversion channel in terms of different gain constant, where the blue, orange, and green correspond to piezo-optomechanical system with thermal bath $n_\text{th}=0,0.5,$ and $1$, respectively; (b) and (c) are the capacity lower bound of induced conversion channel in terms of system cooperativities for $n_\text{th}=0,1$, with each data optimized over the gain constant $\kappa$; (d) The $E_\text{F}$ of the entangled MO state. In (b), (c) and (d), the white dashed line divides the system from stable (lower right) from unstable regime (up left). The red curves separate the parameter regime where direct transduction has zero capacity (lower left) or potentially positive capacity (upper right). In all plots, the extraction ratio $\zeta_\text{o}=0.8,\zeta_\text{e}=1$. 
\label{fig2}}
\end{figure*}

In Ref.~\cite{zhong2020}, we proposed an entanglement-based quantum transduction scheme which first generates high fidelity MO entanglement and then completes the quantum transduction by teleportation. The idea behind is based on a well-known result in quantum channel theory: a very noisy channel can have positive quantum capacity with the assistance of a classical communication channel. In this section, we make this idea more concrete based on the piezo-optomechanical system. We show the system is able to generate MO entanglement which induces an entanglement-based transduction channel with positive quantum capacity even when the system has zero capacity to perform any direct quantum transduction. Instead of using red detuned laser, we pump the optical mode on the blue side band with $\Delta_\mathrm{o}>0$. Adopting the rotating wave approximation, the Hamiltonian looks
\begin{equation}
\begin{split}
    \hat{H}/\hbar=&-\Delta_\text{o}\hat{a}^\dagger\hat{a}+\omega_\text{m}\hat{b}^\dagger\hat{b}+\omega_\text{e}\hat{c}^\dagger\hat{c}+g_\text{om}(\hat{a}^\dagger\hat{b}^\dagger+\hat{a}\hat{b})\\
    &+g_\text{em}(\hat{b}^\dagger\hat{c}+\hat{b}\hat{c}^\dagger),
\end{split}    
\end{equation}
where the optical and mechanical modes are driven in the parametric down conversion regime and two mode squeezed state can be generated. Meanwhile the mechanical excitation can swap to the microwave mode through the piezo-electrical coupling, leading to an entangled MO output state. Ideally, the output entangled state is a two mode squeezed vacuum, while in reality, the thermal noise and dissipation will degrade it to a mixed two mode squeezed Gaussian state. The output state can be obtained in a scattering picture, where the input Gaussian state (vacuum or thermal) is transformed into a Gaussian state under a Gaussian unitary. A Gaussian unitary is equivalently described by a symplectic transformation on the state quadrature \cite{gosson2006}. To obtain this transform, we first write down the Heisenberg-Langevin equation for each mode and combine the input-output relation
\begin{equation}\label{dynaEq}
    \begin{split}
        \dot{\textbf{a}}&=\textbf{M}\textbf{a}+\textbf{N}\textbf{a}_\text{in}\\
        \textbf{a}_\textbf{out}&=\textbf{N}^\text{T}\textbf{a}-\textbf{a}_\text{in}
    \end{split}
\end{equation}
where we group the operators into the following vectors (similar to the previous section) $\mathbf{a}_\text{in}=(\hat{a}^\dagger_\text{in,c},\hat{a}^\dagger_\text{in,i},\hat{b}_\text{in},\hat{c}_\text{in,c},\hat{c}_\text{in,i})^\mathrm{T}$, $\textbf{a}=(\hat{a}^\dagger,\hat{b},\hat{c})^\text{T}$ and $\textbf{a}_\text{out}=(\hat{a}^\dagger_\text{out,c},\hat{a}^\dagger_\text{out,i},\hat{b}_\text{out},\hat{c}_\text{out,c},\hat{c}_\text{out,i})^\text{T}$. The resonance condition is taken ($\Delta_\text{o}=\omega_\text{m}=\omega_\text{e}$). The matrix
\begin{equation}
\mathbf{M}=
\begin{pmatrix}
-\frac{\kappa_\text{o}}{2} & -ig_\text{om} & 0\\
ig_\text{om} & -\frac{\kappa_\text{m}}{2} & ig_\text{em}\\
0 & ig_\text{em} & -\frac{\kappa_\text{e}}{2}
\end{pmatrix},
\end{equation}
\begin{equation}
\textbf{N}=
\begin{pmatrix}
\sqrt{\kappa_\text{o,c}} & \sqrt{\kappa_\text{o,i}} &0 &0& 0\\
0&0&\sqrt{\kappa_\text{m}}&0&0\\
0&0&0&\sqrt{\kappa_\text{e,c}}&\sqrt{\kappa_\text{e,i}}
\end{pmatrix}.
\end{equation}
Taking the mode operators into the frequency domain, we can find $\textbf{a}_\text{out}=\tilde{\textbf{S}}\cdot\textbf{a}_\text{in}$, where $\tilde{\textbf{S}}=\textbf{N}^\text{T}(-i\omega\textbf{I}_3-\textbf{M})^{-1}\textbf{N}-\textbf{I}_5$. Using the relation
\begin{equation}
    \begin{pmatrix}
    \hat{q}\\
    \hat{p}
    \end{pmatrix}=\begin{pmatrix}
    1&1\\-i&i
    \end{pmatrix}\begin{pmatrix}
    \hat{a}\\
    \hat{a}^\dagger
    \end{pmatrix},
\end{equation}
we can convert the scattering matrix into the corresponding quadrature representation
\begin{equation}
    \textbf{x}_\text{out}=\textbf{S}\cdot\textbf{x}_\text{in},
\end{equation}
where $\textbf{S}$ is the desired symplectic transform matrix. The vectors $\textbf{x}_\text{in/out}$ collect all the input and output mode quadratures. If we label the two-mode (microwave and optical) output state quadratures as $\mathbf{x}=\{ \hat{q}_\mathrm{o},\hat{p}_\mathrm{o},\hat{q}_\mathrm{e},\hat{p}_\mathrm{e} \}^\mathrm{T}$, a corresponding covariance matrix $\mathbf{V}_{\mathrm{oe}}^\mathrm{out}$ with the elements defined by $V_{ij}=\frac{1}{2} \left\langle \{ \hat{x}_i-\braket{\hat{x}_i},\hat{x}_j-\braket{\hat{x}_j} \} \right \rangle$ can be obtained, and it can be expressed in the standard form
\begin{equation}\label{vout}
\mathbf{V}_{\mathrm{oe}}
=
\begin{pmatrix}
\mathbf{V}_A & \mathbf{V}_C \\
\mathbf{V}_C^T & \mathbf{V}_B
\end{pmatrix}
% =
% \begin{pmatrix}
% u(\omega) & 0 & w(\omega) & 0\\
% 0 & u(\omega) & 0 & -w(\omega)\\
% w(\omega) &0 & v(\omega) & 0\\
% 0 & -w(\omega) & 0 & v(\omega)
% \end{pmatrix}
\end{equation}
where $\mathbf{V}_A=u(\omega)\textbf{I}_2, \mathbf{V}_C=w(\omega)\textbf{Z}_2,
\mathbf{V}_B=v(\omega)\textbf{I}_2$. This matrix fully characterizes the output MO Gaussian state (ignoring the first moment of each mode since we only care about the state entanglement), where the diagonal elements $u(\omega),v(\omega)$ represent the corresponding output power spectrum densities and the element  $w(\omega)$ indicates the quadrature correlations. 
Again, picking the resonant frequency ($\omega=0$), the matrix elements can be simplified as 
\begin{equation}
    \begin{split}
        &u=1+\frac{8C_\text{om}[1+N_\text{th}+C_\text{em}(1+N_\text{th}-N_\text{th}\zeta_\text{e})]\zeta_\text{o}}{(1-C_\text{om}+C_\text{em})^2}\\
        &v=1+\frac{8[C_\text{em}(C_\text{om}+N_\text{th})-(C_\text{om}-1)^2(\zeta_\text{e}-1)N_\text{th} ]}{(1-C_\text{om}+C_\text{em})^2\zeta^{-1}_\text{e}C^{-1}_\text{om}}\\
        &w=\frac{4[1+C_\text{em}+C_\text{om}+2N_\text{th}C_\text{om}(1-\zeta_\text{e})+2N_\text{th}\zeta_\text{e}] }{(1-C_\text{om}+C_\text{em})^2/\sqrt{C_\text{om}C_\text{em}\zeta_\text{e}\zeta_\text{o}}}.
    \end{split}
\end{equation}
To show this state is indeed entangled, we calculate the two mode Gaussian state entanglement of formation ($E_\text{F}$) \cite{spyros2017} (see the appendix for the definition of entanglement of formation). As shown in Fig.~\ref{fig2}(d), the $E_\text{F}$ is positive at a large parameter regime, providing a good MO entanglement resource for teleportation (details shown in the following section). It is worth mentioning that the system could be unstable since we are using a blue detuned pump. By checking the stability condition \cite{DeJesus87,Yingdan15,Lin13}, we numerically identity the white dashed line separating the stable (lower right) from the unstable regime (upper left corner) in Fig.~\ref{fig2}(b). The intuition is that when the blue-detuned pump becomes too strong, the optomechanical parametric gain will be too large and cause instability.

\subsection{Teleportation induced transduction channel}

With the MO entanglement generated from the piezo-optomechanics, bidirectional quantum transduction can be achieved using teleportation. In this section, we show this entanglement-based conversion induces a Gaussian channel which can reach none zero quantum capacity in large parameter space. Assume we want to convert an input microwave state with covariance matrix $\mathbf{V}_\text{in}$ to the optical regime. According to the standard protocol \cite{furusawa1998,pirandola2006,Pirandola2015}, we first send the input mode and the microwave mode of the entangled source $\mathbf{V}_\text{oe}$ through a $50:50$ beam splitter and perform Homodyne measurement to get $p$ and $q$ quadratures from the two outputs, respectively. Upon a conditional displacement, the input state can be recovered on the optical side.   

The output state can be conveniently derived in the Wigner representation (see appendix for a brief review). Initially, we have a three-mode Wigner function up to normalization
\begin{equation}
    W_i(\mathbf{x})\propto e^{-\frac{1}{2}\mathbf{x}^\mathrm{T}(\mathbf{V}_\text{oe}\oplus\mathbf{V}_\text{in})^{-1}\mathbf{x}},
\end{equation}
where $\mathbf{x}=(\mathbf{x}_\text{o},\mathbf{x}_\text{e},\mathbf{x}_\text{in})$. After the beam splitter, the Homodyne measurement and the feed forward correction, the final Wigner function of the optical mode is given by
\begin{equation}
    W_f(\mathbf{x}_\text{o})\propto\int d\mathbf{x}_\text{in}d\mathbf{x}_\text{e} e^{-\frac{1}{2}\mathbf{x}^\mathrm{T}[\mathbf{F}^\mathrm{T}\mathbf{U}^\mathrm{T}_\text{bs}(\mathbf{V}_\text{oe}\oplus\mathbf{V}_\text{in})^{-1}\mathbf{U}_\text{bs}\mathbf{F} ]\mathbf{x} },
\end{equation}
where $\mathbf{U}_\text{be}$ denotes the beam splitter unitary. The matrix $\mathbf{F}$ corresponds to the displacement operation which takes the form 
\begin{equation}
    \mathbf{F} = 
    \begin{pmatrix}
    \mathbf{I}_2 & \sqrt{2}\mathbf{t}_1 & \sqrt{2}\mathbf{t}_2\\
    0&\mathbf{I}_2&0\\
    0&0&\mathbf{I}_2
    \end{pmatrix}
\end{equation}
where $\mathbf{t}_1=\kappa({\mathbf{I}_2+\mathbf{Z}_2})/{2}$ and $\mathbf{t}_2=\kappa(\mathbf{Z}_2-\mathbf{I}_2)/2$, $\mathbf{Z}_2$ is Pauli-z matrix and $\kappa$ is an arbitrary gain factor. To identify the teleportation induced quantum channel, one can continue evaluating the integral. Instead, to avoid this tedious integral, we go to the characteristic function by Fourier transforming the Wigner function. Remembering a general Gaussian integral formula
\begin{equation}
    \int d\mathbf{x}e^{-\mathbf{x}^T\mathbf{V}\mathbf{x}+\mathbf{x}^T\xi}\propto e^{\frac{1}{4}\xi^T\mathbf{V}^{-1}\xi},
\end{equation}
the output characteristic function can be shown to be specified by the first sub-block of the inverted matrix
\begin{equation}
    [\mathbf{F}^T\mathbf{U}^T_\text{bs}(\mathbf{V}_\text{oe}\oplus\mathbf{V}_\text{in})^{-1}\mathbf{U}_\text{bs}\mathbf{F}]^{-1},
\end{equation}
which corresponds to the output covariance matrix. Straightforwardly, by picking out the first sub-block, we find the input covariance matrix is transformed as $\mathbf{V}_\text{in}\rightarrow\mathbf{T}\mathbf{V}_\text{in}\mathbf{T}^T+\mathbf{N}$ with
\begin{equation}
\begin{split}
\mathbf{T}&=\kappa\mathbf{I}_2,\\
\mathbf{N}&=\mathbf{V}_A-\mathbf{V}_C\mathbf{Z}_2\mathbf{T}-(\mathbf{V}_C\mathbf{Z}_2\mathbf{T})^T+\mathbf{T}^T\mathbf{Z}_2\mathbf{V}_B\mathbf{Z}_2\mathbf{T}\\
&=(v\kappa^2+u-2w\kappa)\mathbf{I}_2.
\end{split}
\end{equation}
Obviously, this defines a single mode Bosonic channel, e.g., when $\kappa<1$, it mimics a thermal loss channel $\mathcal{N}^\prime(\eta^\prime,n_e^\prime)$ with an effective transmissivity $\eta^\prime$ and effective thermal noise 
\begin{equation}
    \eta^\prime={\kappa}^2<1,n_e^\prime=\frac{v\kappa^2+u-2w\kappa}{2\abs{1-\kappa^2}}-\frac{1}{2}.
\end{equation}
Noticing the gain factor $\kappa$ is arbitrary, the effective $\eta^\prime$ thus can be larger than $0.5$, making a positive quantum capacity possible, as detailed in the next section. When the gain factor is chosen $\kappa>1$, it mimics a thermal amplification channel $\mathcal{A}^\prime(\eta^\prime,n_e^\prime)$ with $\eta^\prime>1$ and the thermal noise given by the same expression. Finally, when the modification constant $\kappa=1$, it gives a random displacement channel with noise variance
\begin{equation}
\sigma^2=v+u-2w.
\end{equation}
We denote this channel as $\mathcal{D}^\prime(1,\sigma^2)$. Interestingly, the noise variance expression coincides with the term in Duan criterion $v+u-2w<1$ \cite{duan2000}, which sufficiently identifies the entanglement of a given continuous variable quantum state. Intuitively, the smaller the term $u+v-2w$ is, the more entangled the MO state is, and thus the smaller the noise variance is in the output of the teleportation induced conversion channel.

\subsection{Entanglement-based conversion admits positive capacity with larger parameter space}

Similar to the thermal loss channel, the exact capacities of the thermal amplification channel and the random displacement channel are still not known. To quantify these channels, we use their lower bounds. For thermal amplification channel, its lower bound has a similar form to the thermal loss channel and they can be put in a combined form
\begin{equation}
    Q_\text{LB}^\mathcal{N^\prime,A^\prime}=\max\{0,\log_2(\frac{\kappa^2}{\abs{1-\kappa^2}})-g(n^\prime_e)\}.
\end{equation}
For random displacement channel, a transmission rate can be achieved by GKP code \cite{gkp2001}, which gives a quantum capacity lower bound
\begin{equation}
    Q^{D^\prime}_\text{LB}=\max\{0,\log_2(\frac{2}{e\sigma^2})\}.
\end{equation}
In Fig.~\ref{fig1}(b), to show the sharp contrast to the ideal DQT scheme, we first plot the quantum capacity of the EQT with ideal parameters (unit extraction ratio and zero system noise; note the effective noise is not zero), showing EQT is indeed having positive quantum conversion rate in larger parameter regimes, even at the regime DQT is useless. In Fig.~\ref{fig2}, we plot the quantum capacity lower bound for more practical parameters. For demonstration, the extraction ratios are fixed at $\zeta_\text{o}=0.8,\zeta_\text{e}=1$ and the thermal bath noise is tuned from zero to one. In Fig.~\ref{fig2}(a), we take the cooperativities $C_\text{om}=C_\text{em}=1$ (DQT useless regime), where we see the EQT capacity lower bound could still be positive by tuning the gain constant $\kappa$ and the channel can tolerate around one thermal noise. In Fig.~\ref{fig2}(b)(c), we scan the cooperativities $C_\text{om}$ and $C_\text{em}$ and get the optimized lower bound (picking the optimal $\kappa$). Positive capacity is seen across the red curve to the lower left, where the DQT is impossible to transmit any quantum information. The fact that EQT has a larger parameter space for positive capacity will make the experimental implementation less demanding than DQT, which is quite appealing especially at this early stage of demonstrating quantum transduction. {As the last point, we see from Fig.~\ref{fig2} that the finite $E_\mathrm{F}$ generally indicates a positive capacity of the EQT channel, and $E_\mathrm{F}$ is slightly larger than the capacity lower bound. The intuition is that the capacity lower bound is usually obtained from one shot coherent information \cite{schumacher1996}, which is smaller than $E_\mathrm{F}$.}

As we finish the work in this subsection, we realize a quite recent work \cite{wu2021} studied {the same topic based on a different transducer model}, also {different representation} is used in deriving the quantum channel.

\section{Microwave-microwave entanglement from swapping}

\subsection{Gaussian dynamics for entanglement swapping}

\begin{figure}[t]
\centering
\includegraphics[width=\columnwidth]{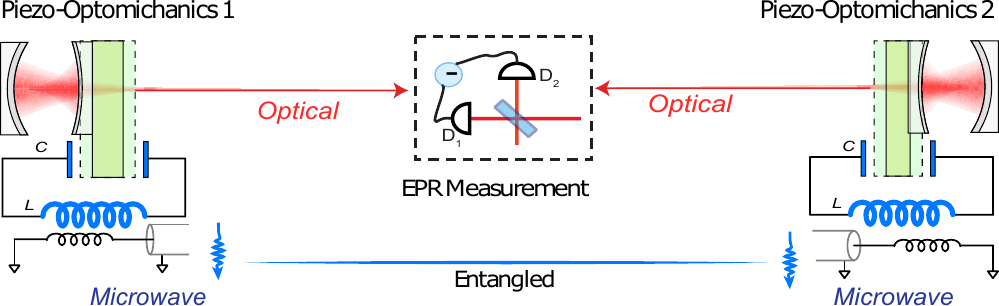}
\caption{Schematic figure for homodyne-based entanglement swapping scheme to generate microwave-microwave entanglement from two piezo-optomechanical systems. The photon click-based scheme can similarly be implemented by replacing the homodyne measurements in the middle by single photon detections.  \label{fig3}}
\end{figure}

The goal of quantum transduction is to connect distant microwave quantum processors. The above schemes achieve this goal by converting the signal to optical regime, transmitting the optical photons through space and converting them back to microwaves. Alternatively, this same goal can be realized if we have faithful MM entanglement, with which we can perform direct microwave signal to signal transmission. Distant MM entanglement can be realized using two piezo-optomechanical systems, as shown in Fig.~\ref{fig3}. The idea is to do \textit{homodyne-based entanglement swapping}---projecting the optical modes onto EPR state---where MM entanglement can be generated. In our setup, we can write down the two MO states as a four mode Gaussian state $\mathbf{V}_\text{oe}^1\oplus\mathbf{V}_\text{oe}^2$,
where $\mathbf{V}^{i=1,2}_\text{oe}$ are given by the output covariance matrix Eq.~\ref{vout}. The optical modes are then sent out for homodyne measurement and the initial entanglement is then expected to be swapped to the microwave modes.

\begin{figure}[t]
\centering
\includegraphics[width=\columnwidth]{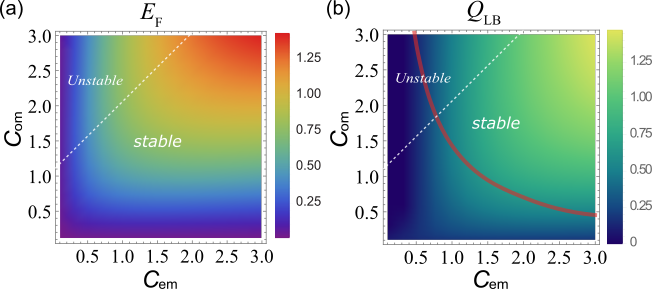}
\caption{(a) The entanglement of formation for the microwave-microwave state after the entanglement swap; (b) The quantum capacity lower bound of the teleportation induced transmission channel using the microwave-microwave entanglement. The white dashed line in each plot separate the piezo-optomechanical system between stable and unstable phase. The red solid line in (b) mark the boundary of parameter regime where direct transduction has zero (lower left) or positive (upper right) capacity in the ideal case. In each plot, we set $\zeta_\text{o}=\zeta_\text{e}=1$ and the thermal bath $n_\text{th}=0$. \label{fig4}}
\end{figure}

To get the expression of microwave entangled state, we briefly review the conditional Gaussian dynamics upon a general-dyne measurement on a portion of a given Gaussian state \cite{alessio2017}. Suppose we have an initial Gaussian state with $n+m$ modes partitioned into $A$ and $B$, respectively. The first moments for each modes are set to be zero and the covariance matrix is
\begin{equation}
    \mathbf{V}=
    \begin{pmatrix}
    \mathbf{\Gamma}_A & \mathbf{\Gamma}_{AB}\\
    \mathbf{\Gamma}_{AB}^\text{T} & \mathbf{\Gamma}_B
    \end{pmatrix}.
\end{equation}
Then we perform a general-dyne measurement on the $m$ modes of the system $B$. Depending on the measurement, one can get conditional state of the system $A$ with $n$ modes. A general-dyne measurement on the $m$ modes is a set of POVM given by 
\begin{equation}
    \hat{F}_{\mathbf{r}_i}=\{\frac{1}{(2\pi)^m}\hat{D}^\dagger_{\mathbf{r}_i}\rho_i\hat{D}_{\mathbf{r}_i}\}
\end{equation}
satisfying $\int_{R^{2m}} d\mathbf{r}_i\hat{F}_{\mathbf{r}_i}=\hat{I}$, where $\mathbf{r}_i\in R^{2m}$ is the measurement outcome and $\rho_i$ is a Gaussian state with zero first moment and second moment $\mathbf{V}_i$. The probability of getting result $\mathbf{r}_i$ is given by
\begin{equation}
    p(\mathbf{r}_i)=\frac{\exp({\mathbf{r}_i^\text{T}\frac{1}{\mathbf{\Gamma}_B+\mathbf{V}_i}\mathbf{r}_i}) }{\pi^m\sqrt{\text{det}(\mathbf{\Gamma}_B+\mathbf{V}_i)}}.
\end{equation}
The state of the $n$ modes of the subsystem $A$ is mapped to \cite{alessio2017}
\begin{equation}
\begin{split}
    \mathbf{V}^i_A&= \mathbf{\Gamma}_A-\mathbf{\Gamma}_{AB}\frac{1}{\mathbf{\Gamma}_B+\mathbf{V}_i}\mathbf{\Gamma}_{AB}^\text{T}\\
    \mathbf{r}^i_A&=\mathbf{\Gamma}_{AB}\frac{1}{\mathbf{\Gamma}_B+\mathbf{V}_i}\mathbf{r}_i
\end{split},
\end{equation}
where we see a remarkable feature of the general-dyne conditioning: the conditional covariance matrix, which determines all correlations, doesn't depend on the measurement outcome.

We now apply this Gaussian conditioning to the entanglement swapping scheme, where we partition the four mode state $\mathbf{V}_\text{oe}^1\oplus\mathbf{V}_\text{oe}^2$ into the optical ($B$) and microwave ($A$) part. With the Eq.~\ref{vout}, we have $\mathbf{\Gamma}_A=v\mathbf{I}_4$, $\mathbf{\Gamma}_B=u\mathbf{I}_4$ and $\mathbf{\Gamma}_{AB}=\text{diag}(w,-w,w,-w)$. The matrix $\mathbf{V}_i$ is first chosen to be a two mode squeezed state
\begin{equation}
    \mathbf{V}_i=
    \begin{pmatrix}
    \cosh(2r)\mathbf{I}_2&\sinh(2r)\mathbf{Z}_2\\
    \sinh(2r)\mathbf{Z}_2 & \cosh(2r)\mathbf{I}_2
    \end{pmatrix},
\end{equation}
then we take the limit $r\rightarrow\infty$ to simulate an ideal measurement. Finally, we obtain a covariance matrix
\begin{equation}\label{mmstate}
    V_\text{MM}=
    \begin{pmatrix}
    (v-\frac{w^2}{2u})\mathbf{I}_2 & \frac{w^2}{2u}\mathbf{Z}_2 \\
    \frac{w^2}{2u}\mathbf{Z}_2 & (v-\frac{w^2}{2u})\mathbf{I_2}
    \end{pmatrix},
\end{equation}
which determines the entanglement of two-mode microwave state.

\subsection{Entanglement for Microwave information transmission}

After the general-dyne measurement, we obtain a microwave two mode Gaussian state given by Eq.~\ref{mmstate}, which is indeed entangled. It can be seen by evaluating the entanglement of formation, as shown in the Fig.~\ref{fig4}(a). This entanglement can be further used as a resource for teleportating quantum information encoded in microwave frequencies, inducing a direct microwave transmission channel. As we did in the previous section, we can similarly evaluate its quantum capacity lower bound, which is shown in Fig.~\ref{fig4}(b). We pick the ideal case with extraction ratio $\zeta_\text{o}=\zeta_\text{e}=1$ and the noise from the thermal bath $n_\text{th}=0$, as compared to the ideal case of direct quantum transduction. We see at the lower left corner delineated by the red solid line, where the direct transduction has zero capacity, the teleportation induced microwave transmission channel still has positive capacity, indicating the advantage of entanglement-based channel in connecting microwave processors.

\begin{table*}[t]
\caption{Comparison of the photon click-based and the homodyne-based entanglement swapping schemes.}\label{tab1}
\begin{center}
\begin{tabular}{c|c}
\hline
\hline
click-based entanglement swapping & homodyne-based entanglement swapping \\
\hline
- in photon number basis & - in continuous variable basis \\
- use single photon detection  & - use homodyne measurement \\
- probabilistic scheme with heralding & - deterministic scheme \\
{- working in low squeezing regime} & {- working in high squeezing regime}\\
- with detection of loss errors & \\
\hline
\hline
\end{tabular}
\end{center}
\label{default}
\end{table*}

\section{Comparison of MM entanglement swapping schemes}

The entanglement swapping can also be discussed in discrete variables similar to the well known DLCZ scheme \cite{duan2001,stefan2021}, where the system typically works at a very different parameter regime, e.g., entangled photon pairs should be generated by the weak parametric down conversion. Also, we need the optical single photon clicks to herald successful MM entanglement, thus this \textit{click-based entanglement swapping} scheme is generally probabilistic. While in the continuous variables, the continuous MM entanglement generation favors the strong parametric down conversion regime and is usually non-probabilistic due to the deterministic property of homodyne measurement \cite{alessio2017}.
At the same time, we expect the MM entanglement from homodyne-based entanglement swapping to be more sensitive to optical photon loss error than that from the photon click-based protocol. To make these comparison more clear, we evaluate the entanglement generation rates from both schemes in the following section. For simplicity, we assume zero intrinsic thermal noises from all modes and the measurement devices are perfect for both protocols.

For the homodyne-based entanglement swapping scheme, the MM entanglement can be quantified by two-mode Gaussian entanglement of formation $E_\mathrm{F}(\omega)$, which measures the amount of entanglement in the output state for a given frequency. In practice, it is important to check the entanglement within certain bandwidth. Because of energy conservation, the overall output state is approximately in a tensor product of all frequency contributions, which indicates that the entanglement is additive. Thus we define a quantity called {entanglement of formation rate} ($E_\mathrm{R}$) as \cite{zhong2020_}
\begin{equation}
E_\text{R}=\frac{1}{2\pi}\int E_\text{F}(\omega)d\omega.
\end{equation}
Intuitively, $E_\text{R}$ tells how efficient a system is in generating entanglement. {Since the entanglement of formation in general upper bounds the distillable entanglement, the rate $E_\mathrm{R}$ actually gives an upper bound of the system entanglement generation rate \cite{guo2019}.} For no optical photon loss, $E_\text{R}$ depends on the MM state as given in Eq.~\ref{mmstate}. If we model an extra optical photon loss as a beam splitter with transmissivity $\tau$, the MM state can be obtained by the replacement $u\rightarrow \tau(u-1)+1$ and $w\rightarrow\sqrt{\tau}w$, which generally results in a reduced entanglement rate.

\begin{figure}[b]
\centering
\includegraphics[width=\columnwidth]{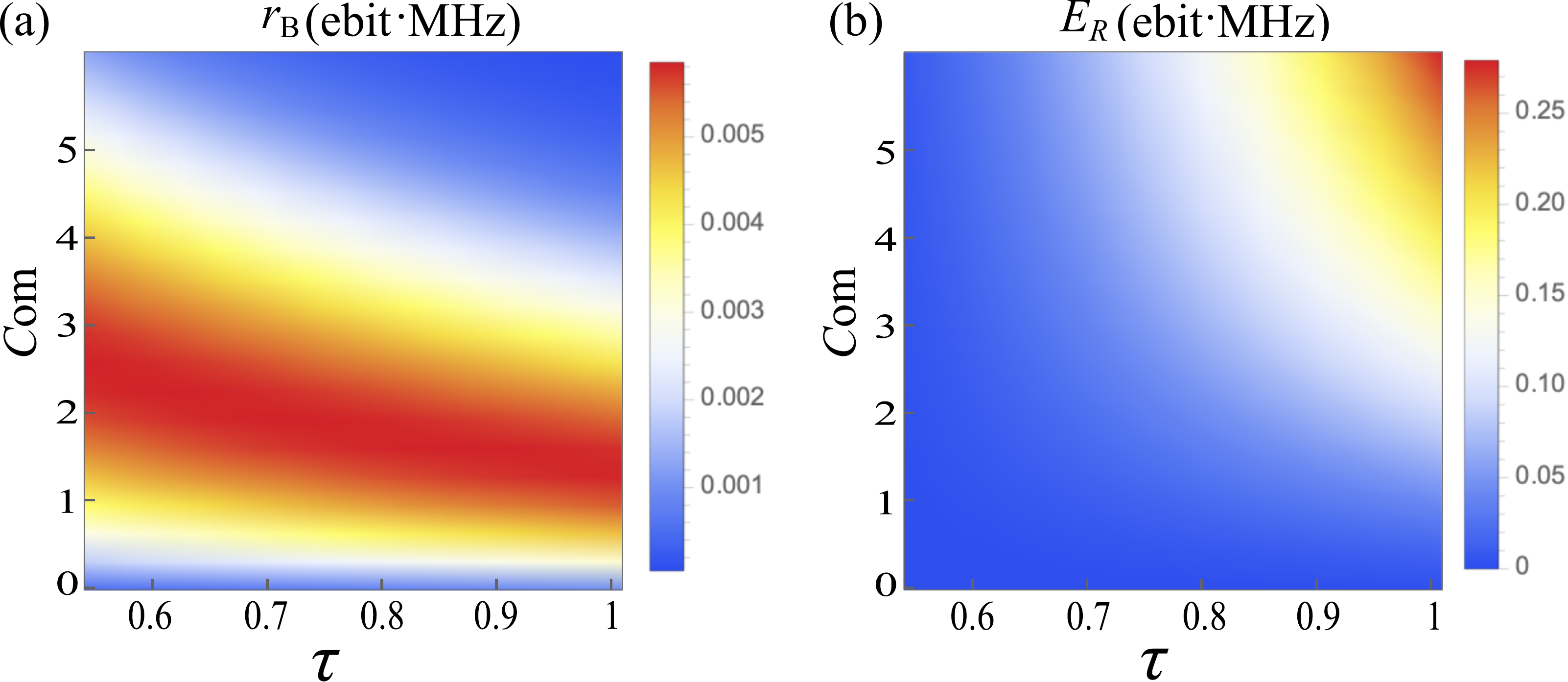}
\caption{(a) The entanglement generation rate for photon click-based scheme
and (b) the entanglement of formation rate for the homodyne-based scheme in terms of $C_\text{om}$ and the optical transmissivity $\tau$. In each plot, we set $C_\text{em}=10$, $\zeta_\text{o}=\zeta_\text{e}=1$ and the thermal bath $n_\text{th}=0$. It is worth mentioning that the non-unit extraction ratios and thermal noise will degrade the entanglement rate for both schemes.
\label{fig5}}
\end{figure}

In the click-based entanglement swapping protocol, optical single photon detection is used to herald the entangled MM Bell pair. The optical photon click can be modeled as a Poisson detection process and the single click probability is given by $r_t\Delta t e^{-r_t\Delta t}$, where $\Delta t$ is pump pulse duration which is typically several microseconds and $r_t$ is the optical photon rate \begin{equation}
    r_t\equiv\braket{\hat{a}^\dagger_\text{out,c}\hat{a}_\text{out,c}}=\frac{1}{2\pi}\int\braket{\hat{a}^\dagger_\text{out,c}[\omega]\hat{a}_\text{out,c}[\omega]}d\omega.
\end{equation} 
Considering there are two piezo-optomechanical devices contributing to the optical heralding event, the single photon click rate can be approximated by
\begin{equation}
    r_B\simeq2r_te^{-r_t\Delta t}
\end{equation}
which is essentially the heralded MM Bell state generation rate.  Similarly, we can also model the optical photon loss by a beam splitter with transmissivity $\tau$, which will reduce the overall Bell pair rate. 

In Fig.~\ref{fig5}, we numerically calculate the entanglement generation rates for both protocols. We see the homodyne-based swapping scheme generally have a larger entanglement rate than the click-based protocol. Meanwhile, as expected the homodyne-based swapping scheme favors the strong parametric down conversion regime (the upper right corner in Fig.~\ref{fig5}(b) where $C_\text{om}$ approaches $C_\text{em}$) and it is very sensitive to the photon loss. In contrast, as shown in Fig.~\ref{fig5}(a), the photon click-based scheme is much more robust to photon loss in generating entanglement. Also, it is preferred for the click-based scheme to work in the weak down conversion regime (smaller $C_\text{om}$) since the single photon pair generation is more probable. In summary, the MM entanglement generation can be done in both discrete and continuous variables and we should properly choose the right protocol according to the practical requirements in the mission of quantum information transduction. For convenience, we summarize their difference in Tab.~\ref{tab1}.

\section{Discussion}

As discussed, EQT is more feasible for quantum state conversion which places much less demanding requirements on the physical implementations. To achieve the EQT, an important step is to successfully demonstrate the MO entanglement. As the experimental technology develops, various physical systems, including piezo-optomechanics, electro-optics, etc., is reported to reach the EQT compatible regime, where high fidelity entanglement could in principle be generated and quantified. In practice, e.g., for discrete variable entanglement, the entanglement verification also needs efficient photon detection. Although optical photon detector can already work with extremely high efficiency, the microwave  detection still suffers from limited sensitivity, thus hindering an efficient MO Bell measurement. Interestingly, with the help of circuit quantum electrodynamics for capturing microwave photons, it is shown that high fidelity single microwave detection is possible \cite{Campagne-Ibarcq2018}. We expect that as the experimental technique improves the MO entanglement should become more controllable, and the EQT would be the first quantum conversion scheme that coherently brings the two important fields---optical communication channel and microwave quantum processor---together; In the mean time, distant MM entanglement from entanglement swapping would become more practical in coherently connecting quantum circuits. In short, these protocols together would bring the ambitious proposal of quantum network more down to earth.

\begin{acknowledgments}
C.Z. thanks Changhun Oh, Yuxin Wang and Filip Rozpedek for helpful discussions. C.Z. and L.J. acknowledge support from the ARO (W911NF-18-1-0020, W911NF-18-1-0212), ARO MURI (W911NF-16-1-0349, W911NF-21-1-0325), AFOSR MURI (FA9550-19-1-0399, FA9550-21-1-0209), AFRL (FA8649-21-P-0781), DoE Q-NEXT, NSF (OMA-1936118, EEC-1941583, OMA-2137642), NTT Research, and the Packard Foundation (2020-71479).
X.H. acknowledges partial support from Argonne National Laboratory Directed Research and Development (LDRD) Program.
This work was performed, in part, at the Center for Nanoscale Materials, a DOE Office of Science User Facility, and supported by the U.S. Department of Energy, Office of Science, Office of Basic Energy Sciences under Contract No. DE-AC02-06CH11357.

\end{acknowledgments}

\begin{appendix}

\section{Characteristic and Wigner function of Gaussian state}

Given any linear operator $\hat{o}\in\mathcal{L}(\mathcal{H})$ acting on the Hilbert space $\mathcal{H}$. One can define the $p$-norm of the operator as $||\hat{o}||_p\equiv[\text{tr}\sqrt{\hat{o}^\dagger\hat{o}}^p]^{1/p}$. When $p =2$, it is called Hilbert-Schmidt norm, based on which one can define Hilbert-Schmidt inner product for two operators
\begin{equation}
    (\hat{o}_i|\hat{o}_j)\equiv{\text{tr}(\hat{o}_i^\dagger \hat{o}_j)}.
\end{equation}
If we confine the linear operator to the Heisenberg-Weyl group elements $\hat{D}_\xi\equiv\exp[-i\bm{\xi}^T\bm{\Omega}\hat{\mathbf{x}}]$ defined on $n$ Bosonic modes, where $\bm{\xi}\in\mathcal{R}^{2n}$, $\hat{\mathbf{x}}=\{\hat{q}_1,\hat{p}_1,...,\hat{q}_n,\hat{p}_n\}^\text{T}$, and the $\Omega$ is the symplectic form
\begin{equation}
    \bm{\Omega}\equiv
    \begin{pmatrix}
    0&1\\
    -1&0
    \end{pmatrix}^{\oplus n},
\end{equation}
we have the orthogonality condition in terms of the Hilbert-Schmidt inner product
\begin{equation}
    (\hat{D}_{\bm{\xi}}|\hat{D}_{\bm{\lambda}})=(2\pi)^n\delta^{2n}(\bm{\xi}-\bm{\lambda}).
\end{equation}
Thus we can view the Weyl operator as defining a set of operator basis, with which any other operator can be expanded. For instance, given a density operator $\rho$, it can be expanded according to
\begin{equation}
    \rho=\frac{1}{(2\pi)^n}\int_{\mathcal{R}^{2n}}d^{2n}\bm{\xi}~\chi(\bm{\xi})\hat{D}_{\bm{\xi}},
\end{equation}
where the expansion coefficient $\chi(\bm{\xi})$ is typically named as the characteristic function, and $\chi(\bm{\xi})=(\hat{D}_{\bm{\xi}}|\rho)=\text{tr}(\hat{D}^\dagger_{\bm{\xi}}\rho)$. Obviously, knowing the characteristic function is equivalent to knowing the state. The Wigner function is defined as the Fourier transform of the characteristic function
\begin{equation}
    W(\mathbf{x})=\frac{1}{(2\pi)^n}\int_{\mathcal{R}^{2n}}d^{2n}\bm{\xi}~\chi({\bm{\xi}})e^{-i\mathbf{x}^\text{T}\bm{\Omega}\bm{\xi}}.
\end{equation}
For Gaussian state with covariance matrix $\mathbf{V}$ and first moment $\bar{\mathbf{x}}$, one can show the characteristic function and the Wigner function take the form
\begin{equation}
    \chi(\bm{\xi})=e^{-\frac{1}{2}\bm{\xi}^T(\mathbf{\Omega}\mathbf{V}\mathbf{\Omega}^T)\bm{\xi}-i(\mathbf{\Omega\bar{x}})^T\bm{\xi}}
\end{equation}
and
\begin{equation}
    W(\mathbf{x})=\frac{e^{-\frac{1}{2}(\mathbf{x}-\bar{\mathbf{x}})^\text{T}\mathbf{V}^{-1}(\mathbf{x}-\bar{\mathbf{x}}) }}{(2\pi)^n\sqrt{\text{det}\mathbf{V}}}.
\end{equation}

\section{Quantum capacity and Gaussian channel}

\subsection{Coherent information}

As capacity for classical channel, quantum capacity is a quantity for measuring the channel's ability to transmit quantum information. In general for many quantum channels, to exactly know the quantum capacity is hard. Instead, lower or upper bound is used for partially describing the channels. In this appendix, we will discuss a lower bound---the coherent information---which defines an achievable rate of a channel to transmit quantum information. 

In general, a quantum channel is defined by a completely positive and trace preserving (CPTP) map (The requirement of a quantum channel to be CPTP is nothing but keeping the quantum process physical)
\begin{equation}
    \mathcal{N}:\rho_A\rightarrow\rho_B,
\end{equation}
where the system input $\rho_A\in\mathcal{H}_A$ and the output $\rho_B\in\mathcal{H}_B$. Theoretically, any quantum channel has a unitary dilation defined as
\begin{equation}
    \mathcal{N}(\rho_A)\equiv\text{tr}_E[\mathcal{U}_{AE}(\rho_A\otimes\ket{0}\bra{0}_E)]
    % \equiv\text{tr}_E(U\rho_A\otimes\ket{0}\bra{0}_EU^\dagger),
\end{equation}
The subscript $E$ is usually to denote the environment input (here we identify it with the output for simplicity). The above dilation naturally defines a complement channel
\begin{equation}
    \mathcal{N}^c(\rho_A)\equiv\text{tr}_A[\mathcal{U}_{AE}(\rho_A\otimes\ket{0}\bra{0}_E)].
\end{equation}
Since the unitary evolution usually correlates the system and the environment, the system output will not contain all the information of the input. The coherent information of a quantum channel is defined as
\begin{equation}
    I_c(\mathcal{N})\equiv\sup_{\rho_A}[S(\mathcal{N}(\rho_A))-S(\mathcal{N}^c(\rho_A))]
\end{equation}
where $S(\rho)\equiv-\text{tr}(\rho\log\rho)$ is the von Neumann entropy. The coherent information has a close connection with conditional entropy, which can be seen by adding an identity channel acting on the purification of the system input. If we denote $\ket{\psi}_{RA}$ as a purification of $\rho_A$, we have a unitary channel $I_R\otimes\mathcal{U}_{AE}$ acting on the input $\ket{\psi}_{RA}\otimes\ket{0}_E$
\begin{equation}
    \rho_{RBE}=I_R\otimes\mathcal{U}_{AE}(\ket{\psi}_{RA}\otimes\ket{0}_E).
\end{equation}
Obviously, the output $\rho_{RBE}$ is a pure state and the coherent information (maximized over the input $\rho_A$) can be written down as
\begin{equation}
    I_c(\mathcal{N})=S(\rho_B)-S(\rho_{BR}),
\end{equation}
which is the negative conditional entropy of the state $\rho_{RB}$. Quantum conditional entropy being negative is a surprising quantum fact compared to classical probability theory, and we see interestingly it defines a lower bound of quantum capacity through the relation with coherent information.

The quantum capacity is defined as the optimal average coherent information when using the channel $n$ times $Q\equiv\sup_n\frac{1}{n}I_c(\mathcal{N}^{\otimes n})$, which in general is difficult to calculate analytically. Since coherent information can be super-additive $I_c(\mathcal{N}_1\otimes\mathcal{N}_2)\ge I_c(\mathcal{N}_1)+I_c(\mathcal{N}_2)$, the single shot evaluation of coherent information usually provides a lower bound of the channel capacity.

\subsection{Gaussian quantum channel}

A Gaussian quantum channel can be specified by its action on the statistical first and second moments of arbitrary Gaussian state $\hat{\rho}(\bar{\mathbf{x}},\mathbf{V})$. In general, we have \cite{eisert2005}
\begin{equation}
\begin{split}
    \bar{\mathbf{x}}&\rightarrow\mathbf{T\bar{x}+d},\\
    \mathbf{V}&\rightarrow\mathbf{TVT^T+N},
\end{split}
\end{equation}
where $\mathbf{T,N}$ are real matrices satisfying the channel completely positive condition
\begin{equation}
    \mathbf{N+i\Omega-iT\Omega T^T}\ge 0.
\end{equation}
$\mathbf{d}$ is usually set to zero since it can be compensated by local displacement and is not affecting the state entanglement. Specifically, when $\mathbf{N}=\mathbf{0}$ and $\mathbf{T}$ is a symplectic matrix, it then defines a Gaussian unitary channel. 

As stated in the main text, the thermal loss channel is modeled as a beam splitter mixing the input mode and the thermal noise
\begin{equation}
    \mathbf{x}\rightarrow\mathbf{\sqrt{\eta}x_\text{in}+\sqrt{\text{1}-\eta}x_{n_e}}.
\end{equation}
For a single mode loss channel $\mathcal{N}(\eta,n_e)$, it is easy to verify that 
\begin{equation}
    \mathbf{T}=\sqrt{\eta}\mathbf{I}_2,\mathbf{N}=(1-\eta)(2n_e+1)\mathbf{I}_2,
\end{equation}
where $\eta<1$ is the transmissivity and $n_e$ denotes the thermal noise.

Similarly, for a single mode thermal amplification channel $\mathcal{A}(\eta,n_e)$: $\mathbf{x}\rightarrow~\mathbf{\sqrt{\eta}x_\text{in}+\sqrt{\eta-\text{1}}x_{n_e}}$ with $\eta>1$, we have
\begin{equation}
    \mathbf{T}=\sqrt{\eta}\mathbf{I}_2,\mathbf{N}=(\eta-1)(2n_e+1)\mathbf{I}_2.
\end{equation}

Random displacement channel $\mathcal{D}(1,\sigma^2)$ can be considered as the limiting case of the above Bosonic channels, where the input signal is contaminated with random Gaussian noise with noise variance $\sigma^2$. We have 
\begin{equation}
    \mathbf{T}=\mathbf{I}_2,\mathbf{N}=\sigma^2\mathbf{I}_2.
\end{equation}

By investigating the coherent information of these Gaussian channels, we can lower bound their quantum capacities. For single mode thermal loss channel $\mathcal{N}(\eta,n_e)$ (or $\mathcal{A}(\eta,n_e)$), the lower bound is given by \cite{weedbrook2021}
\begin{equation}
    I_c(\mathcal{N}(\eta,n_e))=\log_2|\frac{\eta}{1-\eta}|-g(n_e),
\end{equation}
where $g(x)=(x+1)\log_2(x+1)-x\log_2 x$. For random displacement channel, a transmission rate can be achieved by GKP code \cite{gkp2001}, which gives the the quantum capacity lower bound
\begin{equation}
    Q^{D^\prime}_\text{LB}=\max\{0,\log_2(\frac{2}{e\sigma^2})\}.
\end{equation}

\section{Entanglement of formation}

Entanglement of formation ($E_\text{F}$) of a general mixed bipartite state is defined as the infimum of the average von Neumann entropy taken over all its possible pure state decompositions
\begin{equation}
    E_\text{F}=\inf_{p_i,\ket{\psi}}\sum_ip_iE(\ket{\psi}_i).
\end{equation}
It has been proven to be an effective entanglement measure for Gaussian states \cite{spyros2017}. For a general two mode Gaussian state, e.g., the output $\mathbf{V}_\text{oe}$ as specified in the text, a lower bound is given by the formula
\begin{equation}
E_\mathrm{F} = \cosh^2r \, \log_2\! \left(\cosh^2 r \right) - \sinh^2r \, \log_2\! \left(\sinh^2 r \right),
\end{equation}
where $r$ is the minimum amount of anti-squeezing needed to disentangled the state
\begin{equation}
r = \frac{1}{4} \ln \left(\frac{\gamma-\sqrt{\gamma^2-\beta_+\beta_-}}{\beta_-} \right),
\end{equation}
with
\begin{equation}
\begin{split}
\gamma =& 2 \left(\det\mathbf{V}_\mathrm{oe} + 1 \right)-(u(\omega)-v(\omega))^2,\\
\beta_{\pm} =& \det \mathbf{V}_A + \det \mathbf{V}_B-2\det \mathbf{V}_C + 2 u(\omega) v(\omega)\\& + 2 w^2(\omega) \pm 4 w(\omega) (u(\omega)+v(\omega)).
\end{split}
\end{equation}
The lower bound is saturated for a two mode Gaussian state in the standard form encountered in this paper. In the main text, we used the above formula to plot the $E_\text{F}$ with the on resonance frequency (taking $\omega=0$).

\section{Bandwidth limited channel capacity}

 In classical Shannon theory, it is well-known that the finite bandwidth of a transmission line gives a finite rate in the data sampling \cite{shannon1949}, which places a constraint on the capacity rate. Similarly, any practical quantum transducer will have finite bandwidth, limiting the information transmission rate. In the expression Eq.~\ref{Cefficiency} for the conversion efficiency, the on resonance frequency ($\omega=0$) is picked, while in general the conversion efficiency is given by
 \begin{equation}
     \eta(\omega)=\frac{4C_\text{om}C_\text{em}}{\abs{C_\text{om}\alpha+C_\text{em}\beta+\alpha\beta\gamma }^2}\zeta_\text{o}\zeta_\text{e},
 \end{equation}
 where $\alpha=1-\frac{2i\omega}{\kappa_\text{e}}$, $\beta=1-\frac{2i\omega}{\kappa_\text{o}}$ and $\gamma=1-\frac{2i\omega}{\kappa_\text{m}}$. Note this efficiency is defined according to $\eta(\omega)\equiv\abs{\hat{a}_\text{out,c}(\omega)}^2/\abs{\hat{c}_\text{in,c}(\omega)}^2$, where the ratio is between two power spectrum density. Thus the corresponding capacity Eq.~\ref{clower} has a unit $[Q^\mathcal{N}_\text{LB}(\omega)]=$ebit/Sec/Hertz. For a transducer with limited bandwidth, one can define a capacity rate that integrates all frequency contribution 
 \begin{equation}
     Q_\text{LB}\equiv\int d\omega Q_\text{LB}^\mathcal{N}(\omega).
 \end{equation}
This quantity has a unit $[Q_\text{LB}^\mathcal{N}]=$ebit/Sec and obviously it is also a lower bound. Since different quantum transducers generally have quite different transmission bandwidth, the capacity rate $Q_\text{LB}$ defined above will be useful in comparing their different transduction abilities.

\end{appendix}

\bibliographystyle{apsrev4-1}
\bibliography{eb}

This is an expansion of the summary submitted to Optica Quantum 2.0 Conference. The submitted manuscript has been created by UChicago Argonne, LLC, Operator of Argonne National Laboratory (“Argonne”). Argonne, a U.S. Department of Energy Office of Science laboratory, is operated under Contract No. DE-AC02-06CH11357. The U.S. Government retains for itself, and others acting on its behalf, a paid-up nonexclusive, irrevocable worldwide license in said article to reproduce, prepare derivative works, distribute copies to the public, and perform publicly and display publicly, by or on behalf of the Government.  The Department of Energy will provide public access to these results of federally sponsored research in accordance with the DOE Public Access Plan. http://energy.gov/downloads/doe-public-access-plan
\end{document}